\documentclass[12pt]{article}
\usepackage[dvips]{graphicx}

\begin{document}

\title{Various approximations for nucleation
kinetics under smooth external conditions}
\author{Victor Kurasov}

\maketitle

\begin{abstract}
Several simple approximations for the
evolution during the nucleation period
have been presented. All of them have
been compared with precise numerical
solution and the errors
 have been estimated.
 All relative errors are enough
 small.
\end{abstract}

In the process of the first order phase
transition which is
ordinary considered on the
example of condensation, i.e. the formation of
a liquid droplets phase in a mother metastable
vapor phase, a
period of embryos formation
 plays the main role. This period is called  a
  nucleation period. Despite the variety
  of substances one can
  extract some common features in
  evolution during this period.
  These features form the base of the
  theoretical description. The most
  precise solution of this problem was given in
  \cite{PhysRevE94}
  and is based on the universality
  of the nucleation period. But
  this universality disappears when we
  consider more specific
  phenomena and more complicated situations.
  That's why we need
some simple approximations which will allow
us to investigate more
complicate situations.
The goal of this publication is to give
these approximations.
Beside the simple functional dependencies we shall see the
physical reasons which lie in the base of
 these approximations.
This gives us an opportunity to give more
simple model for the
evolution during the nucleation
period which can be used in
consideration of more specific
characteristic of the phase
transition, for example, for
manifestation of stochastic effects
of embryos appearance.

\section{Main equations}

As an example we consider the
case of free molecular regime of
growth of embryos in a three
dimensional space. Then the rate of
growth of the embryo with $\nu$
molecules inside in time $t$
is given by a
standard relation
$$
\frac{d \rho_{dr}}{dt}
=
 \zeta / \tau, \ \ \  \rho_{dr} \equiv \nu^{1/3}
$$
Here
$\tau$ is some characteristic time and we
extract the dependence on
the density of the vapor phase via
supersaturation $\zeta$ which
is
$$
\zeta = \frac{n}{n_{\infty}} - 1
$$
where
$n$ is the vapor molecules density,
$n_{\infty}$ is the same characteristic
for the saturated vapor.

The value of supersaturation $\zeta$
characterizes the power of
metastability in the system.
The analogous value which would be
attained in the system without
embryos formation will be denoted
by $\Phi$. This value is completely
determined by external
conditions and, thus, is supposed to be known.

The distribution of the embryos $f$
over their sizes can be taken as the
stationary one and
the following approximation for $f$
dependence on $\zeta$ can be
written
\cite{PhysRevE94}
$$
f(\zeta) =
f(\Phi_*) \exp(\Gamma \frac{\zeta - \Phi_*}{\Phi_*})
$$
Here and later index $*$
marks  values at some
characteristic moment $t_*$
which can be chosen
as the moment when the half of the total number
of droplets is formed.
The value $\Gamma$ is the derivative of the
critical embryo free
energy over the supersaturation
multiplied by $\Phi_*$.

During the nucleation period one can use
linearization of the ideal supersaturation and
after the appropriate choice of scale
$$
\Phi = \Phi_* + z
$$

Here
$z$ is rescaled time
(see \cite{PhysRevE94}).
One can introduce
$x \sim z - a \rho_{dr}$ ($a$ is  a
constant  \cite{PhysRevE94})
and establish correspondence between
$z$ and $x$.
Under the collective regime of vapor
consumption which takes place
under the free molecular regime
the conservation law for a
substance can be written as \cite{PhysRevE94}
\begin{equation}\label{1}
g = c \int_{-\infty}^{z} (z-x)^3 \exp(x-g(x))
dx
\end{equation}
with the appropriate value of the constant $c=0.189$.
Then the curve $\exp(x-g(x))$ has maximum at $x=0$
(this can be regarded as a choice of
$c$).

Now we shall construct approximations for
solution
$f \sim
\exp(x-g(x)) $
of (\ref{1}). It is shown in Figure 1.
\begin{figure}[hgh]

\includegraphics[angle=270,totalheight=8cm]{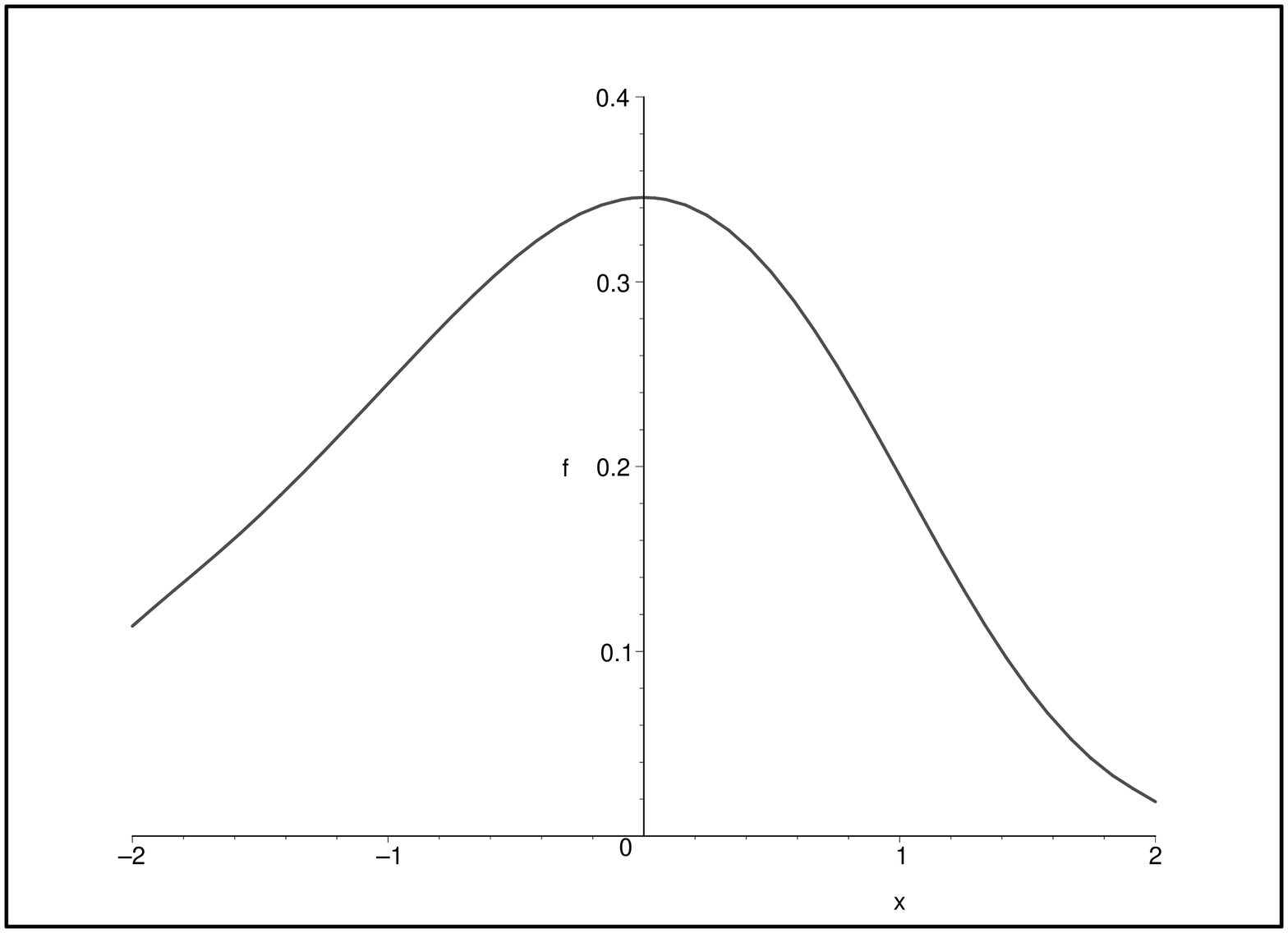}

\begin{caption}
{
Precise solution  for the size spectrum}
\end{caption}

\end{figure}

This curve has a universal form which has been
noticed in \cite{PhysRevE94}.

\section{Iteration approximations}

The iteration solution of eq. (\ref{1}) was given
in\footnote{Solution
in \cite{Kuniprepdyn} requires
the new set of characteristic
parameters at every step
of iteration procedure.}
\cite{Novosib}. The recurrent procedure
is defined as
$$
g_{i+1} = c \int_{-\infty}^{z} (z-x)^3 \exp(x-g_i(x))
dx
$$
with zero approximation $g_0 = 0$.
The first iteration for $f(x) \sim \exp(x-g(x))$
has the following form
$$
f_1 \equiv \exp(x- p c \exp(x))
$$
$$
p=\int_0^{\infty} x^3 \exp(-x) dx = 6
$$
$$
pc = 1.134
$$

One can see that the first iteration is very
close to the real precise solution which can be
seen from Figure 2.
\begin{figure}[hgh]

\includegraphics[angle=270,totalheight=8cm]{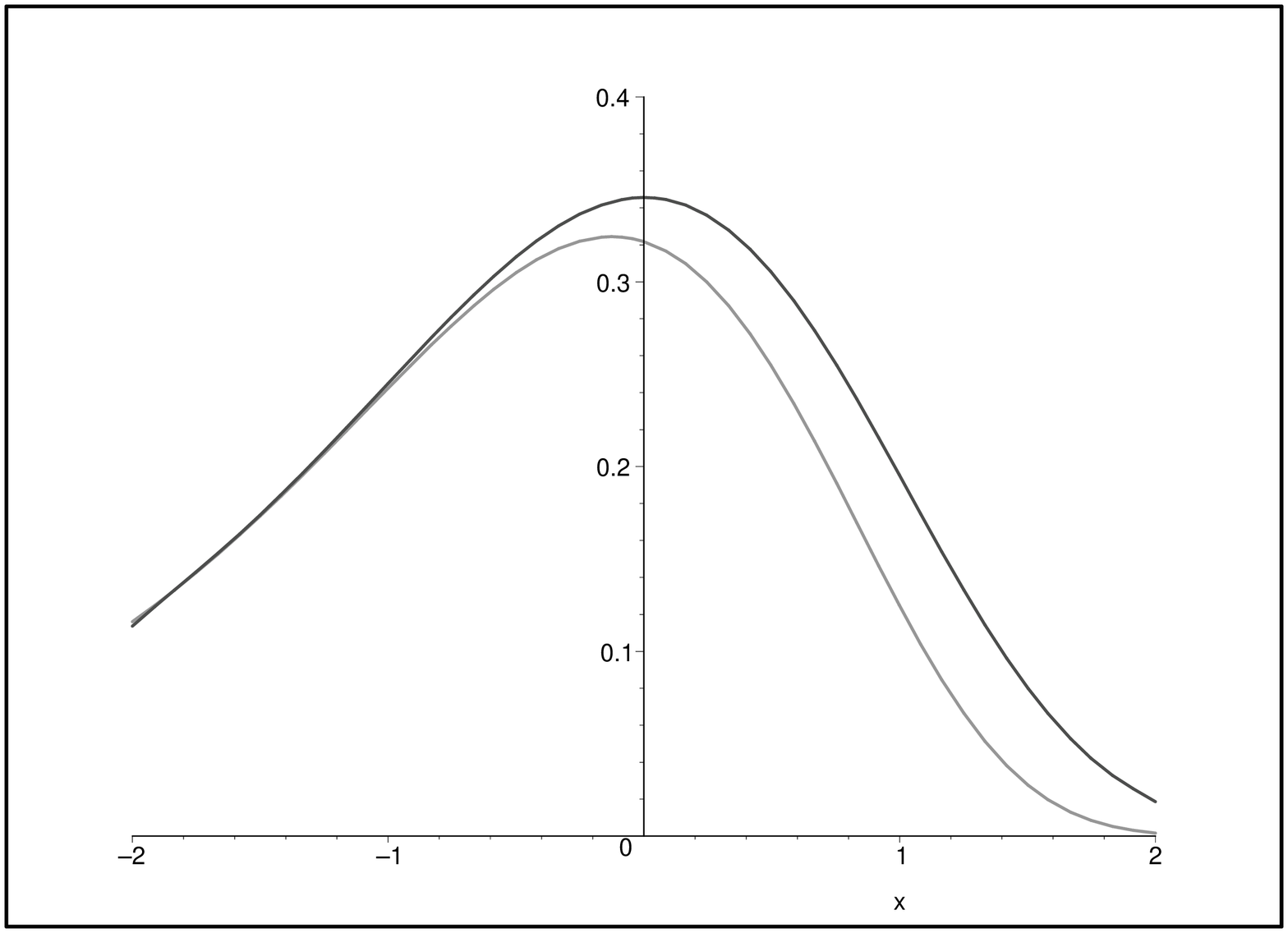}

\begin{caption}
{
Precise solution  and the first iteration
for the size spectrum}
\end{caption}

\end{figure}
The upper curve is the precise solution,
the lower one is the
first iteration.

One can see that the first iteration is already a
good approximation.
Now we shall formulate some new
approximations with more simple physical sense.

\section{Similarity of the embryos formation
conditions }

Under the conditions of the metastable phase
decay one can see in \cite{stoh} the similarity
of conditions under which the embryos are
appearing in the system. It was the consequence
of the scale transformation of time. Here the
situation seems to be another. The time
explicitly stands in the term $\exp(x)$
for $f \sim \exp(x-g(x))$ which
characterizes the action of external forces.
Let us consider the subintegral expression in the
integral calculated for the first iteration. The
size $\rho_{dr}$ of the embryo can be expressed as
$$
\rho_{dr} = \frac{\Phi^2_* }{\Gamma \tau d\Phi / dt}
(z-x)$$
To observe the similarity
we need to see the subintegral equation in
the variable $\rho = z-x$.
The subintegral
function $g_{sub\ 1}$
in the expression
for $g_1$
is the following
$$
g_{sub\ 1} \sim \exp(z)  \rho^3 \exp(-\rho)
$$
Factor $\exp(z)$ is not more than a scaling
factor and we have the universal expression
$$
g_{sub\ 1} \sim   \rho^3 \exp(-\rho)
$$
for subintegral function in the first
iteration. Because  the first iteration resembles
the precise solution we see that the last
relation means that conditions for appearance of
the droplets in different moments of time are
similar, only the scaling factors change. Here it
isn't clear how to use this property, but in the
direct investigation of stochastic effects
 this property  will be very useful.

The universal form of $g_{sub \ 1}$ is drawn in
Figure 3.
\begin{figure}[hgh]

\includegraphics[angle=270,totalheight=8cm]{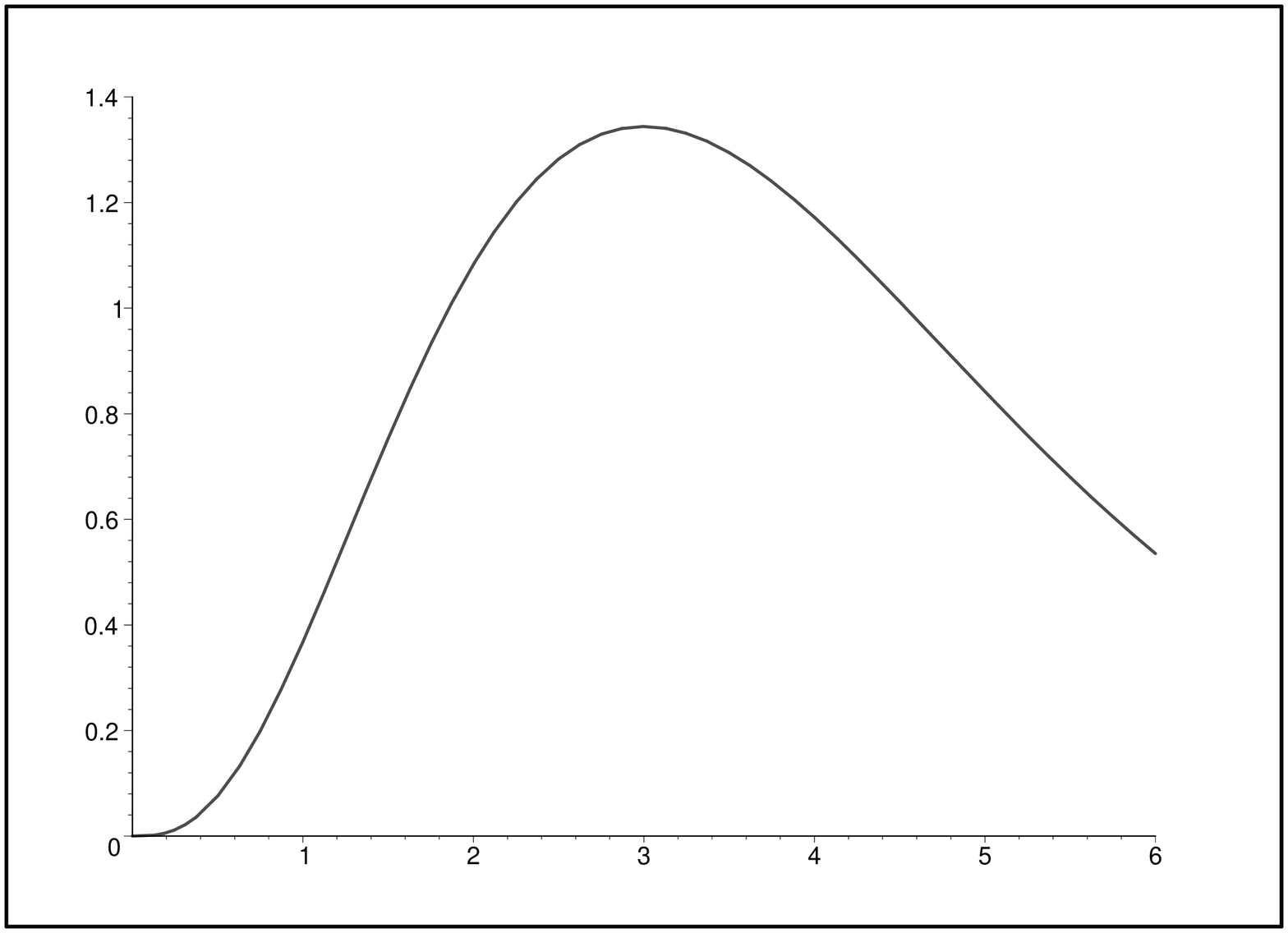}

\begin{caption}
{
The universal form of subintegral expression for
the first iteration
for the size spectrum}
\end{caption}

\end{figure}

We see that the maximum of subintegral expression
is at $\rho = 3$ and at $\rho = 1.4 \equiv \rho_b$
the half of
the maximal height is attained. This value
can be regarded as some characteristic boundary.
The other boundary is
$\rho=\rho_c=5.8$.

\section{First iteration with shifted maximum}

The value $c=0.189$ corresponding to the maximum
at $x=0$  was established after explicit numerical
solution of (\ref{1}). One can pose a question
what value of $c$ corresponds to the maximum of
the first iteration at $x=0$. Certainly,
this procedure will
lead to $c=1/6$ and to a shifted first iteration
$$
f_{1\ sh} \sim \exp(x-\exp(x))
$$
This iteration coincides with precise solution
 even better than the previous first iteration.
 It can be seen from Figure 4.
\begin{figure}[hgh]

\includegraphics[angle=270,totalheight=8cm]{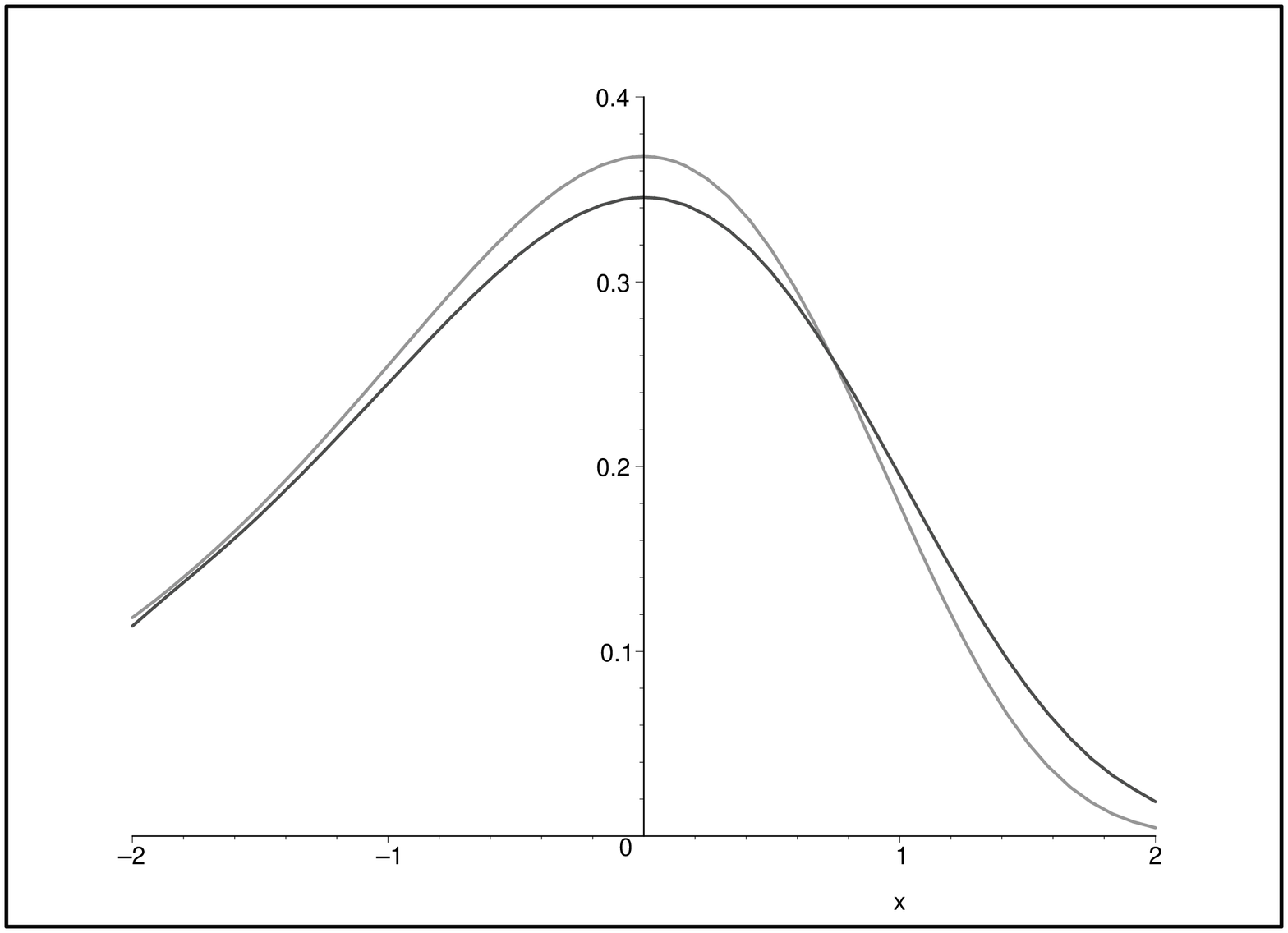}

\begin{caption}
{
The form of precise solution and the shifted
first iteration }
\end{caption}

\end{figure}
There are two curves: the curve with sharp
slope  corresponds to the shifted first iteration
and the curve with smooth slope corresponds to
the
precise solution.

\section{Monodisperse approximation}

Let us calculate expression for $g$
according to (\ref{1}).
We shall use a monodisperse approximation.
We get
$$
g = c \int_{-\infty}^{z} (z-x)^3 \exp(x-g(x)) \approx
c N z_{max}^3
$$
where $z_{max}$ is the coordinate of a
monodisperse peak
and  $N$ is the number of droplets.
For $N$ one can propose
$$
N = \int_{-\infty}^{z} \exp(x-g) dx
$$
For $z_{max}$ since the maximum occurs at $z=-3$
one can get
$$
z_{max} = z+3
$$

This forms the first variant of monodisperse
approximation.
It is not very precise because  we have
attributed  the
value $z+3$ to  small droplets which
are in a big
quantity when we are near maximum of
supersaturation. So,
we have to correct our approximation.

The region of integration to calculate $N$ can
be cut off at $\rho_b$.
Then
the renormalized number of droplets
formed until maximum of supersaturation
is
$$
\hat{N}(z=0)
= \int_{-\infty}^{-\rho_b} \exp(x-g) dx
$$
For
$\rho > \rho_b$ one can
assume that $g$ is negligible.
This leads to the possibility to use
the ideal supersaturation
instead of the real one.
Then one can  come to
$$
N
\approx
 \int_{\rho_b}^{\infty} \exp(-\rho) d\rho
 =
 \exp(-\rho_b) = \exp(-1.4)
$$
So, at every moment the number of
droplets is
$\exp(-1.4) \exp(z)$ and the coordinate
of monodisperse spectrum is $x=3$.
Then for $f \sim \exp(x-g)$ one can get
$$
f \approx f_{mono\ 1} = \exp(x -  3^3 c
\exp(-1.4) \exp(x) )
$$
Figure 5 illustrates this approximation,
here the upper curve is the
precise solution and the lower curve is
the current approximation.
\begin{figure}[hgh]

\includegraphics[angle=270,totalheight=8cm]{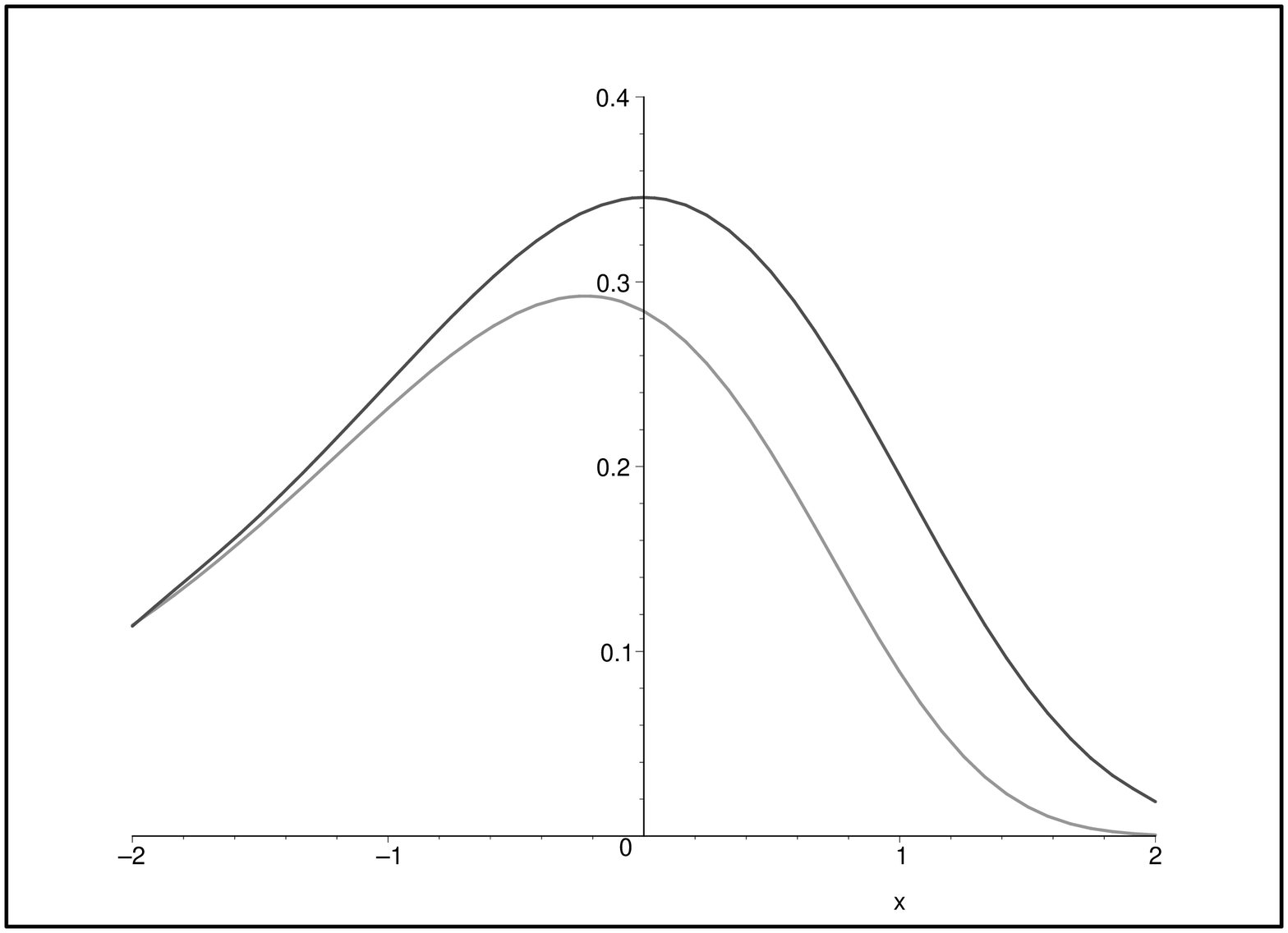}

\begin{caption}
{
The form of monodisperse approximation $f_{mono\ 1}$.}
\end{caption}

\end{figure}

The precision of approximation is satisfactory.

This was only one variant of
monodisperse approximation. Now we shall
present another variant.
Note that at every moment one can
say that the droplet
corresponding to the maximum of subintegral
equation (in the
first iteration) has a size $\rho$ three units
greater than a
current size. This leads to
$$
f \approx f_{mono\ 2} \equiv
\exp(x-(x+3)^3 \exp(-1.4) c)
$$

The form of spectrum is drawn in
Figure 6.
The smooth hill corresponds to the
approximation, the sharp hill
corresponds to the precise numerical solution.
\begin{figure}[hgh]

\includegraphics[angle=270,totalheight=8cm]{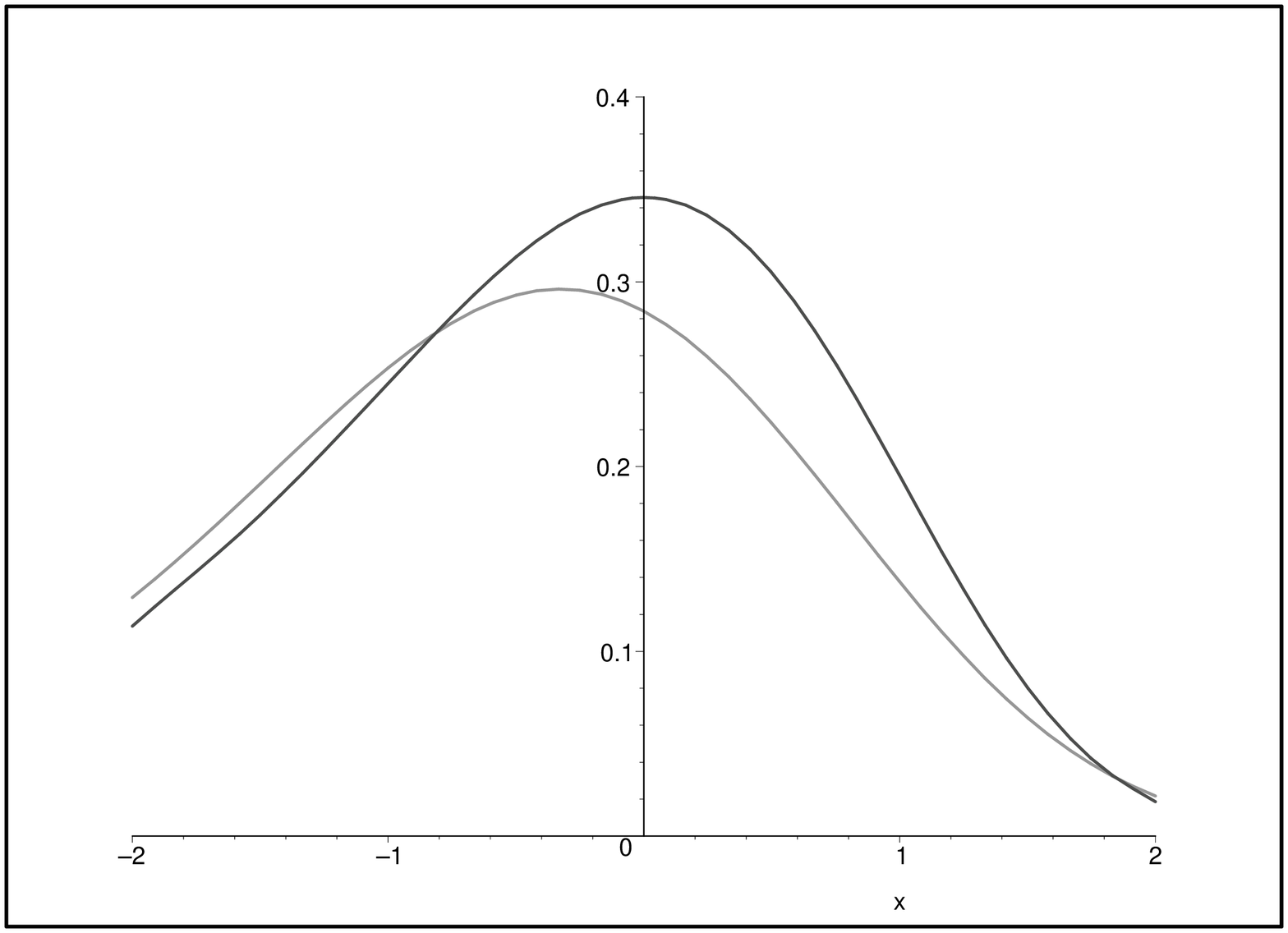}

\begin{caption}
{
The form of monodisperse
approximation $f_{mono\ 2}$  }
\end{caption}

\end{figure}

The precision of the last
approximation is satisfactory.

\section{Monodisperse
approximation at the maximum of
supersaturation}

Earlier we constructed monodisperse
approximations for every
moment of time. Now we shall
take into account that the main
quantity of droplets will be
formed near the maximum of
supersaturation, i.e. near $z=0$.
So, here we can put the position of
monodisperse spectrum as $z=-3$. Then
one can state that the maximum of the
model spectrum
$$
f  \sim
\exp(x - A (x+3)^3)
$$
with parameter $A$  has
maximum at $z=0$. This gives $A = 1/27$.
Moreover one can move back
from the shifted spectrum to the unshifted
one by the substitution of  $ 6 c / 27$
instead of $1/27$.
 Then one can  come to
$$
f \approx f_{mono\ 3} \equiv
\exp(x-\frac{6c}{27} (x+3)^3)
$$
This spectrum together with the
precise
solution is drawn in
Figure
7. Here the sharp curve corresponds to
the precise
solution and
the
smooth curve
corresponds  to the approximation.
It is seen that they are
practically similar.
\begin{figure}[hgh]

\includegraphics[angle=270,totalheight=8cm]{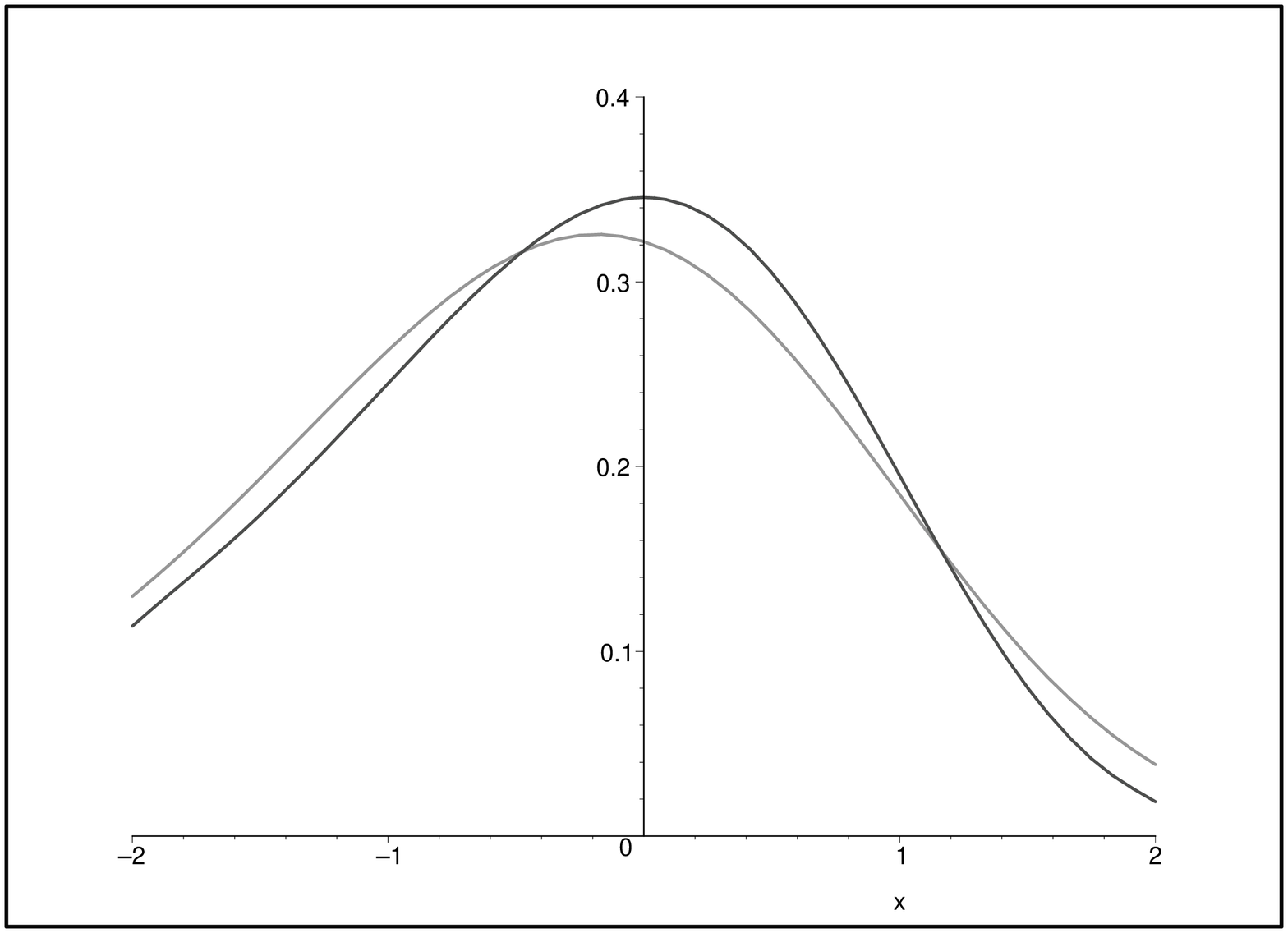}

\begin{caption}
{
The form of monodisperse
approximation $f_{mono\ 3}$  }
\end{caption}

\end{figure}

When $A=1/27$ one can come to
$$
f \approx f_{mono\ 4}
\exp(x-\frac{1}{27}(x+3)^3)
$$
This spectrum together with the
precise solution is drawn in
Figure
8. Here the lower curve corresponds
to precise solution and the
upper curve - to approximation.
It is seen that they are also practically
similar, especially their forms.
\begin{figure}[hgh]

\includegraphics[angle=270,totalheight=8cm]{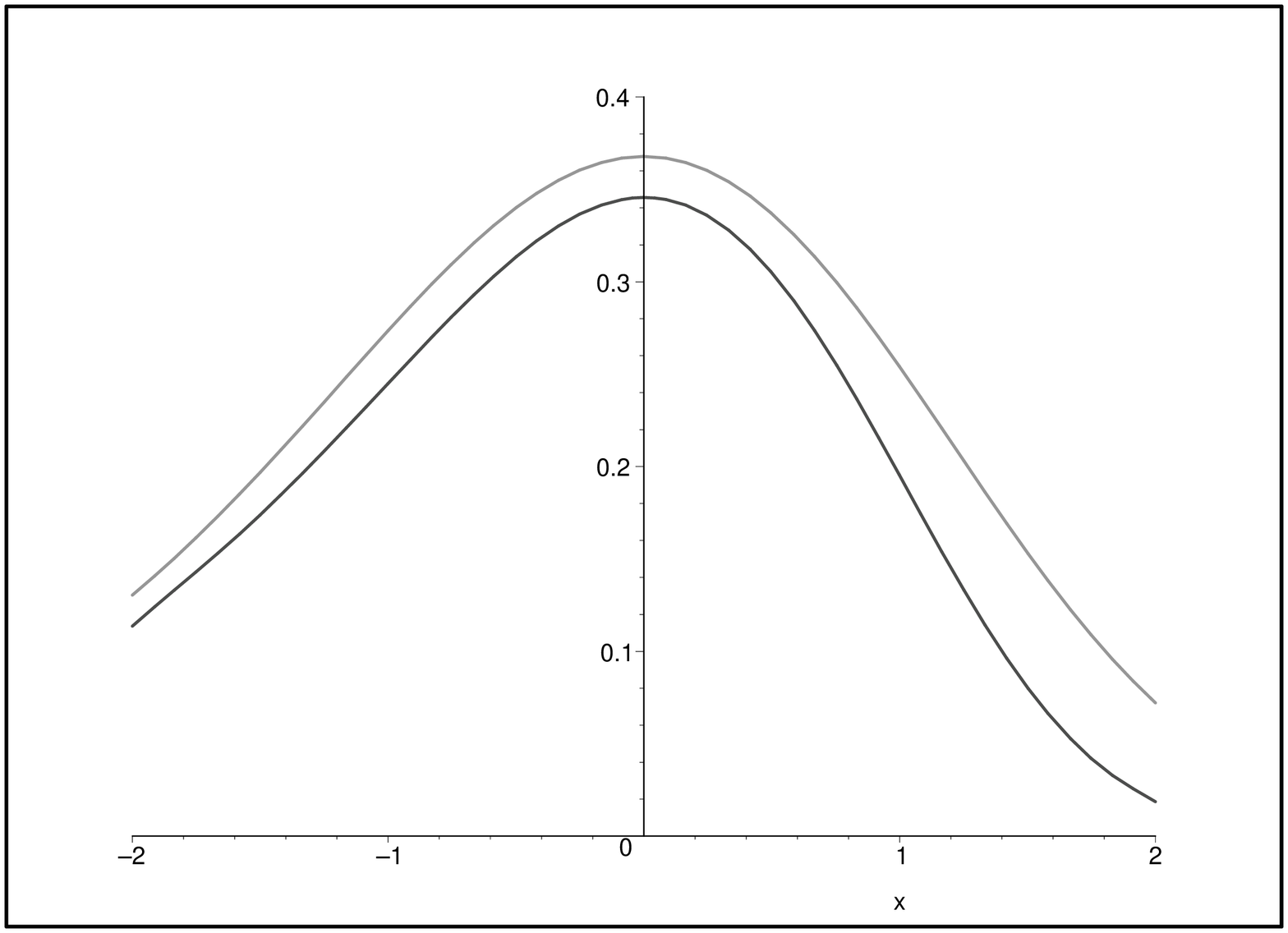}

\begin{caption}
{
The form of monodisperse approximation $f_{mono\  4}$  }
\end{caption}

\end{figure}

Instead of $N \sim \exp(-1.4)\exp(x)$
we put $N\sim 6c/27$ or $N \sim 1/27$
and  conserve the fixed coordinate
$x=3$.

The same can be done with
approximations of the type $f_{mono\ 1}$
and $f_{mono\ 2}$ Then we come to
$$
f_{mono\ 5} = \exp(x-\frac{6c}{27}\exp(x)3^3)
$$
and
$$
f_{mono\ 6} = \exp(x-\frac{1}{27}\exp(x)3^3)
$$
We see that they coincide with $f_{mono\ 3}$
and $f_{mono\ 4}$.
They are drawn in Figure 7 and Figure
 8.
 So, we see that we can do the
 mentioned substitution at different
 steps of derivation.

\section{Two stage scheme}

The next approach to
construct some approximations will be based
on the property which
results in applicability of the first
iteration. What is the
physical reason of applicability of the first
iteration instead of
the precise solution? The
physical reason is the
following: the embryos which
are the main consumers of vapor were
formed at  initial moments
of nucleation period when the
supersaturation was near the
ideal value. This allows us to
extract the initial stage of
nucleation period  $z<  - 1$
and the second stage which is the
whole nucleation period except
the initial stage. The second stage is the
stage when the
main quantity of embryos was formed.

Really $g(-1) <  c \exp(-1)$ is
small in comparison with
characteristic value of $g$ during
the second stage $g \sim c$.
So at $z< - 1$ the
supersaturation $\zeta$ is close to $\Phi$.

The influence of  embryos appeared
at initial stage on the evolution at the
second stage is given through
$$
a_i
=
\int_{-\infty}^{-1}
x^i \exp(x-cg(x))
dx
\ \ i=0,1,2,3
$$

Direct calculation on a base of
the first iteration leads to
$$
a_0 = \exp(-1)
$$
$$
a_1 = -2 \exp(-1)
$$
$$
a_2 = 5 \exp(-1)
$$
$$
a_3 = - 16 \exp(-1)
$$
Then
$$
g \approx g_{in} \equiv c(x^3 a_0-3x^2a_1+3xa_2-a_3)
$$
and the spectrum $f \sim \exp(x-g)
= \exp(x - c(x^3 a_0-3x^2a_1+3xa_2-a_3))$
is drawn in Figure 9 together with
the precise solution
which has more
sharp form.
\begin{figure}[hgh]

\includegraphics[angle=270,totalheight=8cm]{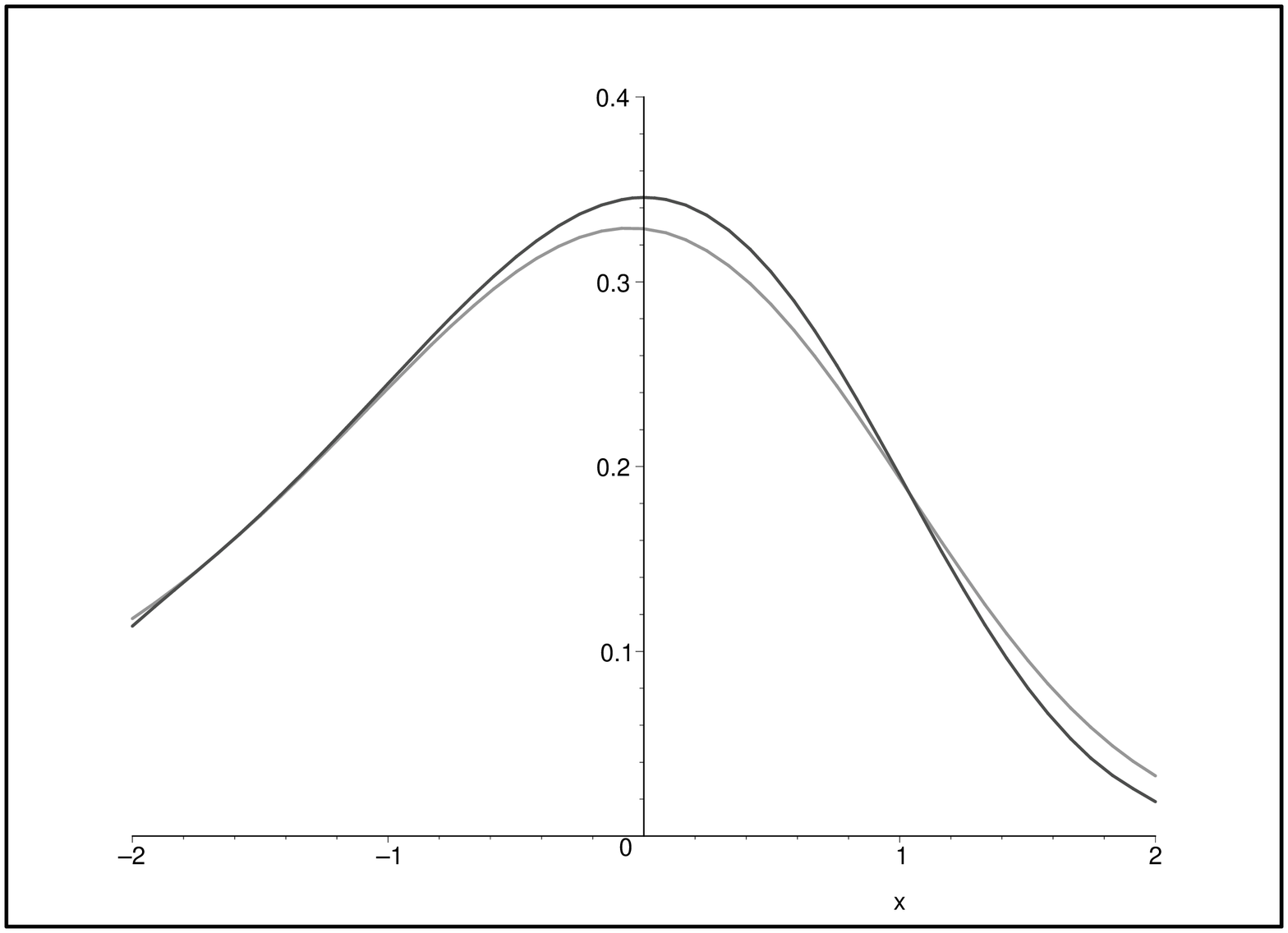}

\begin{caption}
{
The form of  approximation of two stages together
with precise solution.  }
\end{caption}

\end{figure}

Here we forget that this expression
takes place only for $z>-1$.
We spread it for $z<-1$ because here
the spectrum isn't too high.

However, the asymptotic behaviour
at $z \rightarrow -\infty $ will be
qualitatively wrong.
But as for the form of spectrum
in a really significant region
we see that the coincidence of two
curves is practically ideal.

\section{Balance property}

The model forms of the size spectrum obtained
above have to be self
consistent. The self consistency can be
treated at least in two
senses. The first is the weak dependence
of the total number of
droplets on the
number of droplets in the
monodisperse peak. This
question is separately analyzed
directly in investigation of
stochastic effects. Here we shall see the
weak dependence of the
total number of
droplets on the position of the monodisperse peak
or on the position of the boundary between
two stages of the
nucleation period.

We shall analyze two models -
the model of the monodisperse peak
and the model appeared in the two stage scheme.
These models are
the most precise ones.

We shall start with the model of the monodisperse
peak. At first
we shall generalize it. Let the monodisperse
peak be formed at
$z=-p$.  Then instead of $A=1/27$ we
have $A=1/(3p^2)$. For the
size spectrum we get
$$
f \sim \exp(x - \frac{1}{3 p^2} (x+p)^3)
$$
The total number of droplets we get the
following  expression
$$
N_{tot} = N_{in} + \exp(-p)
$$
where $N_{in}$ is the number of
droplets appeared after $z=-p$. The
last value can be calculated as
$$
N_{in} =
\int_{-p}^{\infty}
\exp(x - \frac{1}{3 p^2} (x+p)^3)
dx
$$

In the renormalization adopted here the
total number
calculated on the
base of precise solution is precisely
equal to $1.00$.
The
numbers $N_{in}$ and $N_{tot}$ are
drawn in Figure 10
as functions
of $p$. It is clear that the dependence
of $N_{tot}$ on $p$ is very
weak one.

\begin{figure}[hgh]

\includegraphics[angle=270,totalheight=8cm]{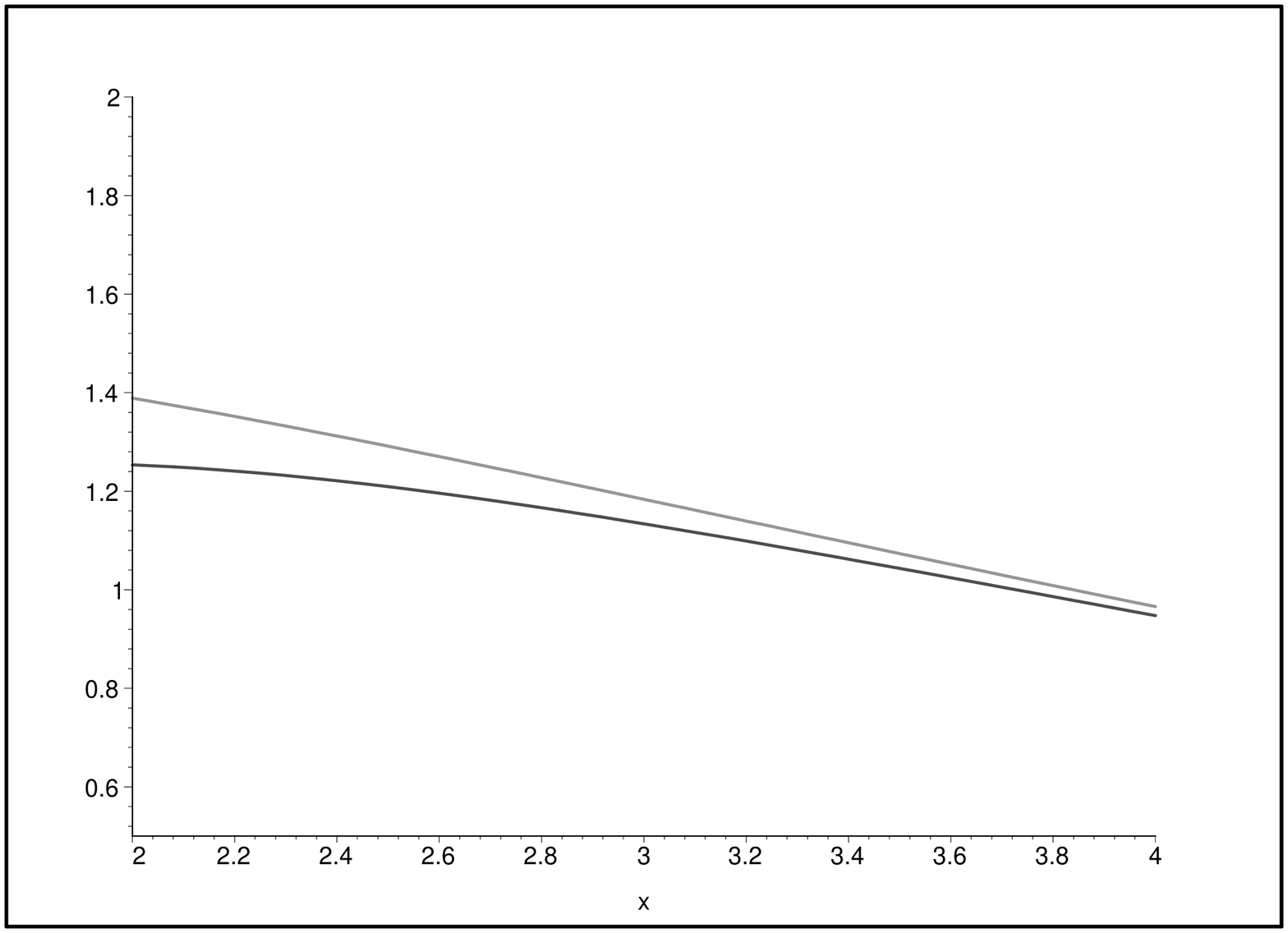}

\begin{caption}
{
The values  $N_{tot}$
and $N_{in}$ as functions of the position of
monodisperse
peak $p$.  }
\end{caption}

\end{figure}

The same we shall do for the model of the
two stage approach. We
shall state that before the
boundary $z=-p$ the spectrum is the
ideal one
$$
f \sim \exp(x)
$$
Fot $z$ greater than $-p$ the evolution is
governed only by the
substance accumulated
in the droplets appeared before $z=-p$.
This
gives
$$
f \sim \exp(x - c(x^3 a_0 - 3 x^2 a_1
+ 3  x a_2 - a_3 ))
$$
where universal momentums are given by
$$
a_0 = \exp(-p)
$$
$$
a_1 =  - p \exp(-p)
$$
$$
a_2 = p^2 \exp(-p) + 2 p \exp(-p) + 2 \exp(-p)
$$
$$
a_3 = - p^3 \exp(-p) - 3 p^2 \exp(-p) - 6 p \exp(-p) - 6 \exp(-p)
$$

The total numbers of
droplets $N_{tot}$ is equal to
$$
N_{tot} = \exp(-p) + N_{in}
$$
where $N_{in}$ is the number of
droplets appeared after $z=-p$.
The last value is given by
$$
N_{in} = \int_{-p}^{\infty}
\exp(x - c(x^3 a_0 - 3 x^2 a_1 + 3  x a_2 - a_3 ))
dx
$$

The
numbers $N_{in}$ and $N_{tot}$ are drawn in Figure
11 as functions
of parameter $p$ of the boundary $z=-p$.
One can see that the dependence
of $N_{tot}$ on $p$ is very weak
one. Moreover the number of
droplets (which is always greater than
the precise value $1.00$) has
minimum at $p \sim 1$. Namely this
value is the optimal one. We
see that the value $p=0$ used in
\cite{stohgrin} is very far from the optimal one.

One can also note that this
property of the weak dependence in
deeply  associated with
constructions in the base of the modified
gaussian method
considered in \cite{Gaussian}.

\begin{figure}[hgh]

\includegraphics[angle=270,totalheight=8cm]{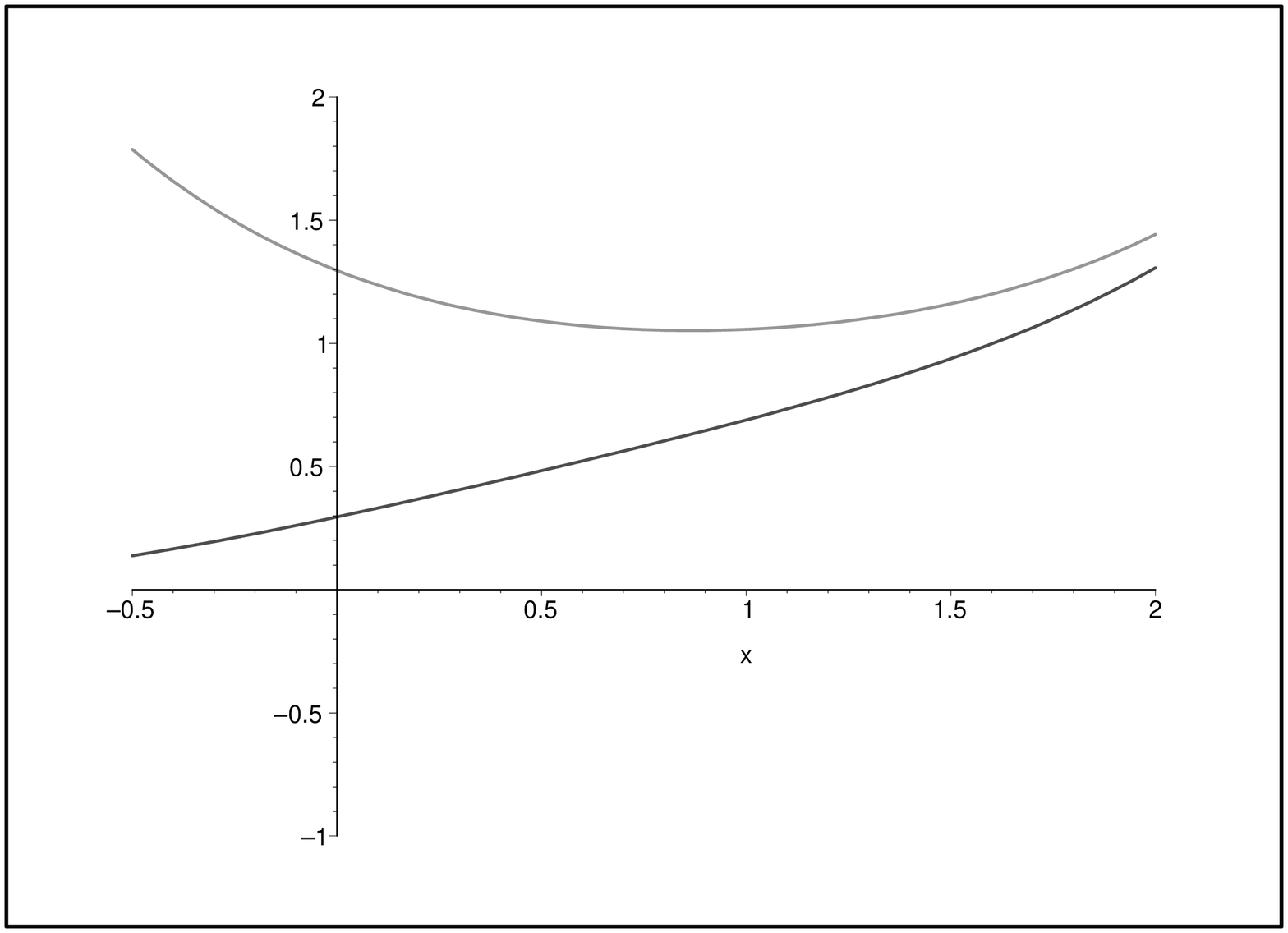}

\begin{caption}
{
The values  $N_{tot}$
and $N_{in}$ as functions of the position of the boundary  $p$.  }
\end{caption}

\end{figure}

\end{document}